\documentstyle[amssymb,preprint,aps]{revtex}
\begin{document}
\title{Frequency Hopping in Quantum Interferometry:\\
Efficient Up-Down Conversion for Qubits and Ebits}
\author{P. Mataloni, G. Giorgi and F. De Martini}
\address{Dipartimento di Fisica and \\
Istituto Nazionale per la Fisica della Materia\\
Universit\`{a} di Roma ''La Sapienza'', Roma, 00185 - Italy}
\maketitle

\begin{abstract}
A novel single-photon Mach-Zehnder interferometer terminated at two
different frequencies realizes the nonlinear frequency conversion of optical
quantum superposition states. The {\it information-preserving} character of
the relevant unitary transformation has been experimentally demonstrated for
input {\it qubits} and {\it ebits}. Besides its own intrinsic fundamental
interest, the new scheme will find important applications in modern quantum
information technology. \ PACS: 03.65.Ta, 03.67.-a, 42.50.-p, 89.70.+c
\end{abstract}


Interferometry of quantum particles is rooted at the core of modern physics
as it provides a unique tool of investigation and a direct demonstration of
fundamental properties of nature as complementarity, nonlocality and quantum
nonseparability~\cite{1}. In modern times all methods and protocols of
quantum information and quantum computation involve interferometry through
the very definition of the conceptual cornerstones of this science, viz. the 
{\it qubit} and the {\it ebit} \cite{2}. In this framework it is well known
that the $1^{st}-order$ interference of particles, e.g. optical photons in
the present work, is a utterly fragile property that can be easily spoiled
by de-coherence when the overall system exhibits a certain degree of
complexity. More fundamentally it is a common notion that the interfering
particles cannot be substantially disturbed by collisions let alone by the 
{\it hardest} possible collisions, i.e. the ones implying the {\it %
annihilation} or the \ {\it creation} of the particles themselves.

In the present paper we demonstrate experimentally that the last seemingly
obvious condition is not a necessary one. In facts, there exits a class of 
{\it information-preserving} unitary transformations of parametric type
allowing a nonlinear (NL)\ frequency conversion of all quantum
interferometric dynamical structures via a\ QED particle annihilation or/and
creation process. Conceptually the overall process may be considered as the
dynamical ''reversal'' of the {\it quantum injected} NL\ parametric particle
amplification/squeezing process that has been realized recently (QIOPA) \cite
{3}. Furthermore the frequency conversion process may be noiseless and may
be easily realized with a ''quantum efficiency'' close to its maximum value, 
$QE$ $=1$. These are indeed very useful properties that will be of large
technological interest in the domain of quantum information and computation.

Refer to Figure 1 showing the schematic diagram of\ a new kind of {\it %
single-photon} Mach-Zehnder (MZ) $1st-order$ interferometer (IF). The
optical structure consists of an input $50/50$ beam splitter (BS)\ coupled
to two photon wavevector (wv) modes ${\bf k}_{j}$ $(j=1,2)$ at the
wavelength (wl) $\lambda $, in our case lying in the infrared (IR) spectral
region. Before standard mode recombination by the output BS$_{2}$, this
simple mode structure is interrupted by a device {\bf U} providing a unitary
nonlinear (NL) transformation ${\bf U}$\ on the input single-photon
superposition state, i.e. the {\it qubit }defined in a 2-dimensional Hilbert 
${\bf k}$ - space{\it :}

\begin{equation}
\left| \Phi \right\rangle =\alpha \left| 1\right\rangle _{1}\left|
0\right\rangle _{2}+\beta \left| 0\right\rangle _{1}\left| 1\right\rangle
_{2}
\end{equation}
where the state labels{\it \ }${\it 1}${\it , }${\it 2}${\it \ }refer to the
IF\ modes $k_{1}$and $k_{2\text{ }}$respectively. Assume now for simplicity
and with no lack of generality: $\beta =\alpha \exp (i\Phi )$ and $\alpha
\equiv 2^{-%
%TCIMACRO{\UNICODE[m]{0xbd}}%
%BeginExpansion
{\frac12}%
%EndExpansion
}$. Let us consider the frequency up-conversion process. The NL
frequency-conversion unitary evolution operator,\newline
\begin{equation}
{\bf U\equiv \exp [}\widetilde{{\bf g}}%
%TCIMACRO{\tsum}%
%BeginExpansion
\mathop{\textstyle\sum}%
%EndExpansion
\limits_{j=1,2}{\bf (}\widehat{{\bf a}}_{j}\widehat{\overline{{\bf a}}}%
_{j}^{\dagger })+h.c.]\newline
\end{equation}
provides the QED\ {\it annihilation} of the input qubit (1) defined by the
momenta $\hslash {\bf k}_{j}$, phase $\Phi $ and wl $\lambda $, and the
simultaneous QED {\it creation} of a new qubit $\left| \Psi \right\rangle $
defined by the momenta $\hslash \overline{{\bf k}}_{j}$ $(j=1,2)$, phase $%
\Psi $ and wl $\overline{\lambda }=(\lambda ^{-1}+\lambda _{p}^{-1})^{-1}$
in our case lying in the ultraviolet (UV) spectral region. The $2^{nd}-order$
tensor parameter $\widetilde{{\bf g}}$ is proportional to the\ interaction
time $t$ , to the $2^{nd}$- order susceptibiliy $d^{(2)}$ of the NL\ medium
and to the ''pump'' field ${\bf E}_{p}$, with wl $\lambda _{p}$, wv ${\bf k}%
_{p}$, phase $\Theta $. The pump field is assumed to be a single plane-wave
coherent ''classical field'' undepleted by the interaction. The wl's and
wv's\ $\overline{{\bf k}}_{j}$, ${\bf k}_{j}$, ${\bf k}_{p}$ are mutally
connected by the energy conservation $\overline{\lambda }^{-1}=(\lambda
^{-1}+$ $\lambda _{p}^{-1})$ and by \ the {\it phase-matching condition} \
(PMC)\ for the 3-wave interaction: ${\bf k}_{j}+\overline{{\bf k}}_{j}+{\bf k%
}_{p}=0$, leading in our simple plane case to an equation with two
solutions: $j=1,2$. Similarly, the qubit phases at different wl's are also
connected by: $\Phi -\Psi +\Theta =0$. Assume $\Theta =0$ for convenience.
We see that by introduction of \ the device U\ and of an additional output $%
BS_{3}$ the standard MZ-IF is transformed into a new kind of interferometer
terminated at two {\it different }wavelengths $\lambda $ and $\overline{%
\lambda }$. Correspondingly two sets of \ {\it interference fringes} can be
retrieved upon changes $\Delta \Phi =$ $2^{3/2}\pi X/\lambda $ of the mutual
phase of the modes ${\bf k}_{j}$ via displacements $X$ of an optical mirror
M\ activated by a piezo-transducer. Note that since in the present
experiment PMC couples {\it deterministically }each mode ${\bf k}_{j}$ to a
corresponding $\overline{{\bf k}}_{j}$, no additional quantum interference
effects arise in the overall NL\ coupling process \cite{4}.

Let us now venture in a more detailed account of the experiment shown in
Figure 2. A cw single-mode, linearly polarized diode laser, Mod. RLT8810MG,
operating at the IR wl $\lambda =876.1nm$ with output power of $2mW$, was
highly attenuated by a set of neutral density filters (ND) to the single
photon level. The achievement of this relevant field's property was tested
by two different Hanbury-Brown-Twiss (HBT) methods A and B. The first method
(A)\ consisted of the well known linear photodetection correlation technique
at the output of a $50/50$ BS. The exact single-photon condition implying
perfect anti-correlation is expressed by the zero value of the degreee of $%
2^{nd}-order$ coherence determined either at wl's $\lambda $ or at wl $%
\overline{\lambda }$: $g^{(2)}(0)=0$ \ \cite{5}. To carry out this
experiment for the wl $\lambda $ (or, independently for wl $\overline{%
\lambda }$) the IF\ phases were set by the mirror M to the value:$\;\Phi $ = 
$\Psi $ = $\pi /2$. This condition, carefully tested in the multiphoton
regime, implied equal photodetection rates at the output of the detectors $%
D_{A}$ and $D_{B}$ for wl $\lambda $ (or $\overline{D}_{A}$ and $\overline{D}%
_{B}\;$for wl $\overline{\lambda }$). In the single-photon regime, i.e.
after beam attenuation, we obtained for the test at wl $\lambda $ a number
of detected concidences $N_{C}=0$\ \ with a number of singles $N_{A}=1015$
and $N_{B}=1223$, over a statistical sample of $10^{5}$ events. This sets
the limit value: $g^{(2)}(0)<8.05\cdot 10^{-2}$. For the test at wl $%
\overline{\lambda }$ the corresponding figures were: $N_{C}=0$, $N_{%
\overline{A}}=810$, $N_{\overline{B}}=830$, $N=10^{5}$ leading to:$\;%
\overline{g}^{(2)}(0)<14.8\cdot 10^{-2}$.

By a second {\it nonlinear} method (B), never adopted previously to our
knowledge, a new $2^{nd}-order$ quantum correlation function was
investigated involving both IR\ and UV\ wavelengths $\lambda $ {\it and} $%
\overline{\lambda }$ and the density operator $\rho $ of the overall field
emerging from the NL\ crystal: 
\begin{equation}
g_{j}^{NL(2)}(0)=Tr(\rho E_{j\lambda }^{(-)}E_{j\overline{\lambda }%
}^{(-)}E_{j\overline{\lambda }}^{(+)}E_{j\lambda }^{(+)})/[Tr(\rho
E_{j\lambda }^{(-)}E_{j\lambda }^{(+)})Tr(\rho E_{j\overline{\lambda }%
}^{(-)}E_{j\overline{\lambda }}^{(+)})]
\end{equation}
\newline
In this case the {\it nonlinear}\ anticorrelation condition $%
g_{j}^{NL(2)}(0)=0$ implied that any up-converted single photon on mode $%
\overline{{\bf k}}_{j}$ was associated with the vacuum field on ${\bf k}_{j}$
and viceversa. In particular it also implied that Fock states $\left|
n\right\rangle $ with $n>1$ were absent on mode ${\bf k}_{j}$ before the
interaction, thus providing a further verification of the expression (1).
This peculiar nonlinear HBT experiment applied to each mode pair $({\bf k}%
_{j},$ $\overline{{\bf k}}_{j})\;j=1,2$ implied the determination of
counting correlations between the output signals generated by the pairs of \
IR\ and UV\ detectors $D_{j}$ and $\overline{D}_{j}$, coupled respectively
to the modes ${\bf k}_{j}$ and $\overline{{\bf k}}_{j}$ {\it after} their
mutual NL interaction. For modes $({\bf k}_{1},$ $\overline{{\bf k}}_{1})$
the following results were obtained:\ coincidences:$\;N_{C}=0$, singles:$%
\;N_{D}=2636$, $N_{\overline{D}}=713$, number of trials: $N=10^{5}$. Thid
led to the upper limit:$\;g_{1}^{NL(2)}(0)<5.3\cdot 10^{-2}$.\newline
All detectors\ operating at the IR wl $\lambda $ were equal Si avalanche
single-photon SPCM-200PQ diodes with Quantum-Efficiency $qe\approx 30\%$
while the two\ detectors operating at the UV\ wl $\overline{\lambda }$ were
photomultipliers:\ Philips-56DUVP $(qe\approx 23\%)$ and Hamamatsu-R943-02 $%
(qe\approx 21\%)$. A computer interfaced Stanford Research 400 counter was
adopted for counting and averaging the detected signals.

The input single-photon field with wl $\lambda $ was injected into the input 
$50/50$ beam splitter $BS_{1}$ of the double MZ-IF\ with output modes ${\bf k%
}_{j}$, $j=1,2$. These modes were mutually $\Phi -dephased$ by a
piezoelectrically driven mirror (M) and the associated fields were brought
by a $f=3.5cm$ lens (L) into a common focal region, with diameter $\varphi
\approx 20\mu m$, within a\ NL $LiIO_{3}$, $l=1mm$ thick crystal slab, cut
for Type I phase matching. Here a strong NL 3-wave interaction took place
between the input field associated with ${\bf k}_{j}$, the up-converted
field associated with the $\overline{{\bf k}}_{j}$ and a single mode high
intensity ''pump'' field associated with the ultrashort pulses emitted with
wl $\lambda _{p}=795nm$ by a mode-locked $76MHz$ Coherent MIRA 900 Ti-Sa
femtosecond laser. Finally, the output beams ${\bf k}_{j}$ emerging from the
NL crystal were again superimposed by a $50/50$ beam splitter BS$_{2}$ thus
completing the usual MZ-IF scheme at the input wl $\lambda $. In a similar
way the two up-converted UV output beams at the UV wl $\overline{\lambda }=$ 
$416.8nm$ were superimposed on an independent $50/50$ BS$_{3}$, thus
completing the MZ-IF scheme at the up-converted wl $\overline{\lambda }$.
The difficult task of filtering the very weak beam at wl $\overline{\lambda }
$ against the very strong UV\ beam al wl $\lambda _{p}/2$ was overcome by
spatial discrimination after the NL crystal and by the adoption of two
interference filters at $416.8nm$ with bandwidth $10nm$.

Note that the up-conversion unitary transformation of the quantum
superposition state (or ''qubit'') at the IR\ wl $\lambda $ into the ''UV\
qubit'' with wl $\overline{\lambda }$ is a {\it noise free }process since
energy conservation doesn't allow any amplification of the input vacuum
state. Of course, the\ inverse transformation process is also possible{\it \ 
}as an input qubit with wl $\overline{\lambda }>\lambda _{p}$ can be
frequency ''down-converted'' into a corresponding one with wl $\lambda =%
\overline{\lambda }\lambda _{p}(\overline{\lambda }-\lambda _{p})^{-1}>%
\overline{\lambda }$. This ''optical parametric amplifying'' (OPA) process
is nevertheless affected by a {\it squeezed vacuum} noise due to the
amplification of the input vacuum state at wl $\lambda $ \cite{3}.

The expression of the quantum efficiency (QE)\ of the up-conversion process,
defined as the ratio between the average numbers of scattered and input
photons on the modes $\overline{{\bf k}}_{j}$, ${\bf k}_{j}$ as a function
of the peak intensity of the pump pulse $I_{p}$ and for perfect collinear
interaction, can be obtained by previous evaluation of the field at the
output of the nonlinear interaction: $U^{\dagger }\widehat{{\bf a}}U$ and $%
U^{\dagger }\widehat{\overline{{\bf a}}}U$ \cite{6}. This leads to the
expression, 
\begin{equation}
QE=\sin ^{2}\left( \frac{\pi }{\varepsilon _{0}}\sqrt{\frac{I_{p}}{\lambda
\lambda _{p}n^{o}\overline{n}^{e}\left( \vartheta _{m}\right) }}d^{\left(
2\right) }l\right) \text{.}
\end{equation}
where $n^{o}$ is the ''ordinary ray'' refraction index at wl $\lambda $
while $\overline{n}^{e}\left( \vartheta _{m}\right) $\ is the
''extraordinary ray '' refraction index at wl $\overline{\lambda }$ for the
angle $\vartheta _{m}$ determined by ${\bf k}_{p}$ and by the optic axis of
the crystal (phase matching angle) \cite{7}.

The expression of $QE$ just given in Equation 4 refers to the case of a
collinear interaction, a condition which can be approximated by improvement
of the spatial superposition of the beams in the NL interaction region. We
report in Figure 3 the theoretical value of $QE$ as function of the pump
intensity $I_{p}$, calculated for the $1mm$ thick $LiIO_{3}$ crystal adopted
in the experiment. The theoretical value of $I_{p}$ corresponding to the
limit condition $QE=1$ is found $I_{p}=200GW/cm^{2}$. This figure may be
compared with the experimental result $QE\approx 0.4$ we obtained in a side
experiment by focusing a low repetition rate $100fs$, $I_{p}=200GW/cm^{2}$
pulse on the same crystal:\ Figure 3 \cite{6}. This discrepancy is
attributed to the non perfect realization of the plane wave collinear
interaction in the active focal spot of the focusing lens. In the present
MZ-IF\ experiment the peak intensity of each Ti-Sa laser pulse was $\approx $
$1GW/cm^{2}$, and the the corresponding measured value of the quantum
efficiency was: $QE\simeq 3\cdot 10^{-3}$.

The persistence of the quantum superposition condition within the $%
IR\rightarrow UV$ frequency hopping process is demonstrated in Figure 4 by
the two correlated interference fringe patterns showing an equal periodicity
upon changes of the mutal dephasing $\Delta \Phi $ of the IR\ modes ${\bf k}%
_{1},$ ${\bf k}_{2}$. As previously emphasized in a different context \cite
{3}, this is but one aspect of a very general {\it information preserving} \
transformation of all unitary NL parametric up- (or down-) conversion
transformations. \ By these ones any input qubit at wl $\lambda $ and
expressed by Equation 1 is generally transformed into another at wl $%
\overline{\lambda }\neq \lambda $ \ keeping the {\it same }complex
parameters $\alpha ,\beta $ of the original one, i.e. fully reproducing its 
{\it quantum information} content. In addition and most important, \ the
present work shows that these transformations can be {\it noise-free} and
can be realized with a quantum efficiency close to its maximum value.

Apart from the fundamental relevance of these results due to the peculiar
paradigmatic and historical status of single-particle interferometry in the
quantum mechanical context, the present work is expected to have a large
impact on modern quantum information technology. This can be illustrated by
the following example. Consider a case in which quantum information is
encoded on a single microwave photon with wl $\lambda $, e.g. within the
cavity of a micromaser. If we want to transfer conveniently this information
at a large distance we need to use an optical fiber exhibiting its low loss
behaviour in the IR spectral region, at wl $\lambda ^{\prime }$. This can be
done in a ''information lossless'' manner in a NL waveguide by \ the
up-conversion $\lambda \rightarrow \lambda ^{\prime }$.\ If \ now this
information is to be transferred to a set of \ trapped atoms in an optical
cavity we may need a further lossless up-conversion into the visible: $%
\lambda ^{\prime }\rightarrow \lambda ^{\prime \prime }$, etc. This scenario
may generally represent an appealing alternative to other linear methods,
but in some cases it may indicate the \ {\it only available}\ \ solution to
sort quantum information out of a nanostructure quantum device and, most
important, to interconnect it efficiently within a large information network
made of heterogeneous components. This may the case of \ a NMR\ quantum gate
operating at a radiofrequency wl \cite{8} or of a\ superconducting quantum
dot gate or a SQUID device operating at still lower electromagnetic (e.m.)\
frequencies \cite{9}.

So far we have been dealing only with conversion of ''qubits''. The
extension of our NL\ method to a two photon entangled state, or specifically
to elementary entangled information carriers, i.e. {\it ebits}, can be
easily realized in several ways depending on the nature of the entanglement.
Consider for instance a linear-polarization $({\bf \pi })$ entangled
2-photon state emitted over two spatial modes ${\bf k}_{1},$ ${\bf k}_{2}$
by a Spontaneous Parametric Down Conversion (SPDC)\ process in a NL\
crystal: $\left| \Phi \right\rangle =\alpha \left| \uparrow \right\rangle
_{1}\left| \uparrow \right\rangle _{2}+\beta \left| \downarrow \right\rangle
_{1}\left| \downarrow \right\rangle _{2}$ \cite{3,10}. With reference to a
technique recently adopted by P.Kwiat {\it et al.} \cite{10} the modes ${\bf %
k}_{j}$ and the strong coherent pump beam with wv ${\bf k}_{p}$ could be
focused by the common lens\ L beam into a combination of \ two equal thin
Type I NL\ plane crystal slabs, e.g. $LiIO_{3}$, $l=1mm$ thick, and placed
in mutual contact along their plane orthogonal to ${\bf k}_{p}$. If \ these
slabs are mutually rotated around the axis parallel to ${\bf k}_{p}$by an
angle $\phi =\pi /2$ and the linear polarization ${\bf \pi }_{p}$ of the
pump beam is also rotated by $\phi =\pi /4$, both nonlocally correlated
orthogonal $\pi -$state components of the injected entangled state undergo
equal NL\ transformations given by [2] thus realizing an overall, {\it %
information preserving} up- (or down-) frequency conversion of \ $\left|
\Phi \right\rangle $.

By recent works\ \cite{11}\cite{12} a new conceptual and formal perspective
has been introduced in quantum information according to which the optical 
{\it \ field's modes } rather than the photons are taken as the carriers of
quantum information and entanglement. Furthermore in that picture any {\it %
qubit} is physically implemented by a two-dimensional subspace of Fock
states of the e.m. field, specifically the state spanned by the vacuum state
and the 1-photon state. According to this perspective the class of
information preserving NL\ transformations of the state given by Equation
[1] investigated in the present work should be more correctly referred to 
{\it entangled states} and may indeed provide a useful new set of unitary
transformations for the Hilbert space evolution of these new information
states. \newline
\ \ \ \ This work has been supported by the FET European Network on Quantum
Information and Communication (Contract IST-2000-29681: ATESIT) and by
PAIS-INFM\ 2002 (QEUPCO).

\centerline{\bf Figure Captions}

\vskip 8mm

\parindent=0pt

\parskip=3mm

Figure 1: Schematic diagram of\ the nonlinear Mach-Zehnder interferometer
(MZ-IF)\ terminated at two different correlated frequencies.

Figure 2: Lay-out of the single-particle MZ-IF experiment.

Figure 3: Up-conversion quantum efficiency $QE$ as function of the laser
pump intensity $I_{p}$ :\ theoretical (continuous line) end experimental
results.

Figure 4: Single photon interference fringes obtained within the same
experiment at the different correlated wavelengths $\lambda $ and $\overline{%
\lambda }=(\lambda ^{-1}+$ $\lambda _{p}^{-1})^{-1}$ . The phase period of
the fringing patterns has been found is in agreement with the figure of
merit $(0.7nm/V)$ of the piezoelectrical transducer activating the mirror M.


\begin{references}
\bibitem{1}  W.K. Wootters and W.H. Zurek, {\it Phys. Rev. D} {\bf 19}
(1979) 473; L.S. Bartell, {\it ibid.} {\bf 21} (1980) 1698; M. Jammer, {\it %
The Philosophy of Quantum}{\ }{\it Mechanics} (Wiley, New York) 1974.

\bibitem{2}  S. Popescu in {\it Introduction to Quantum Computation and
Information}, H. Lo, S.Popescu and T. Spiller Eds.(World Scientific, N.Y.)
1998

\bibitem{3}  F. De Martini, Phys. Rev.Lett.{\bf 81} (1998) 2842 and:\
Phys.Lett.A {\bf 250, }15 (1998), F.De Martini, V. Mussi and F. Bovino,
Optics Comm.{\bf 179, }581{\bf \ }(2000), F. De Martini, G. Di Giuseppe and
S. Padua, Phys.Rev. Lett. 87, 150401 (2001).

\bibitem{4}  It is possible to conceive a NL\ 3-wave interaction in which
the PMC\ is relaxed, e.g. in a very thin crystal, and the input modes are
coupled {\it non-deterministically} to the up-converted ones. The Feynman
path indistinguishability implied by the additional quantum chance condition
may lead to additional interference phenomena that interplay with the
dominant behaviour of the nonlinear\ MZ-IF.

\bibitem{5}  R. Loudon, {\it The Quantum Theory of Light}:\ Ch.5 (Clarendon
Press, Oxford 1983) .

\bibitem{6}  P.Mataloni, O. Jedrkiewicz and F.\ De Martini, Phys. Lett. A 
{\bf 243,} 270 (1998).

\bibitem{7}  R. W. Boyd, {\it Nonlinear Optics}: Ch. 2${\Bbb \ }$(Academic
Press, San Diego,1992) .

\bibitem{8}  I. L. Chuang in {\it Introduction to Quantum Computation and
Information}, H. Lo, S.Popescu and T. Spiller Eds.(World Scientific, N.Y.)
1998.

\bibitem{9}  Y. Nakamura, Y. Pashkin and J.S.Tsai, Nature {\bf 398}, 786
(1999); D. Loss and D. Divincenzo, Phys.Rev.A {\bf 57}, 120 (1998).

\bibitem{10}  P. G. Kwiat, E. Waks, A. G. White, I. Appelbaum and P.H.
Eberhard, Phys. Rev. A {\bf 60} R773, (1999).

\bibitem{11}  E. Knill, R. Laflamme and G. Milburn, {\it Nature} (London)
396, 52 (1998). E. Lombardi, F. Sciarrino, S. Popescu and F. De Martini,
quant-ph/0109160 and: Phys.Rev.Lett. 88, 18/2, (2002).

\bibitem{12}  M. Duan, M. Lukin , J. Cirac and P. Zoller, {\it Nature}
(London) 414, 413 (2001); F. Sciarrino, E. Lombardi and F. De Martini,
quant-ph/0201019 v1; and submitted for publication.\newpage
\end{references}
\end{document}